%% file: main.tex
\documentclass[manuscript, screen, nonacm, dvipsnames]{acmart}

\usepackage{fancyhdr}
\usepackage[normalem]{ulem}
\usepackage{algorithmic}
\usepackage{graphicx}
\usepackage{textcomp}
\usepackage{tcolorbox}
\usepackage{siunitx}
\usepackage{listings}
\usepackage{tabularx}
\usepackage{fancyvrb}
\usepackage[labelfont=bf, textfont=bf, aboveskip=5pt, belowskip=0pt, parskip=0pt]{caption,subfig}
\usepackage{enumitem}
\usepackage[italic]{mathastext}

\newif\ifcameraready
\camerareadytrue
\definecolor{JungleGreen}{RGB}{41,171,135}
\newcommand{\insitu}{\textit{in situ}}
\newcommand{\vpass}{\textit{$V_\text{pass}$}}
\newcommand{\vread}{\textit{$V_\text{read}$}}
\newcommand{\vprog}{\textit{$V_\text{program}$}}
\newcommand{\verase}{\textit{$V_\text{erase}$}}
\newcommand{\vth}{\textit{$V_\text{th}$}}

\usepackage{tikz}
\newcommand{\ballnumber}[1]{\tikz[baseline=(myanchor.base)] \node[circle,fill=.,inner sep=1pt] (myanchor) {\color{-.}\bfseries\footnotesize #1};}

\ifcameraready
  \newcommand{\todo}[1][]{}
  \newcommand{\sg}[1]{{#1}}
  \newcommand{\rw}[1]{{#1}}
  \newcommand{\ben}[1]{{#1}}
  \newcommand{\revSG}[1]{\textcolor{black}{#1}} %
  \newcommand{\revRW}[1]{\textcolor{black}{#1}} %
  
  \newcommand{\urevRW}[1]{\textcolor{black}{#1}} %
  \newcommand{\urevSG}[1]{\textcolor{black}{#1}} %
  \newcommand{\tRW}[1]{\textcolor{black}{#1}} %
  \newcommand{\tSG}[1]{\textcolor{black}{#1}} %
\else
  \newcommand{\todo}[1][]{\textbf{\scriptsize \fcolorbox{black}{red}{\color{white}{TODO}}} \underline{$\overline{\hbox{\emph{#1}}}$}}
  \newcommand{\sg}[1]{\textcolor{black}{#1}}
  \newcommand{\rw}[1]{\textcolor{black}{#1}}
  \newcommand{\ben}[1]{\textcolor{black}{#1}}
  \newcommand{\revSG}[1]{\textcolor{black}{#1}} %
  \newcommand{\revRW}[1]{\textcolor{black}{#1}}
  
  \newcommand{\urevRW}[1]{\textcolor{black}{#1}} %
  \newcommand{\urevSG}[1]{\textcolor{black}{#1}} %
  \newcommand{\tRW}[1]{\textcolor{magenta}{#1}} %
  \newcommand{\tSG}[1]{\textcolor{Green}{#1}} %
\fi

\DeclareSIUnit{\nothing}{\relax}

\AtBeginDocument{%
  \providecommand\BibTeX{{%
    \normalfont B\kern-0.5em{\scshape i\kern-0.25em b}\kern-0.8em\TeX}}}

\setcopyright{None}
\copyrightyear{2024}

\begin{document}

\title{TCAM-SSD: A Framework for Search-Based Computing in Solid-State Drives}

\author{Ryan Wong}
\affiliation{
    \institution{University of Illinois Urbana-Champaign}
    \country{USA}
}
\author{Nikita Kim}
\affiliation{
    \institution{Carnegie Mellon University}
    \country{USA}
}
\author{Kevin Higgs}
\affiliation{
    \institution{University of Illinois Urbana-Champaign}
    \country{USA}
}
\author{Engin Ipek}
\affiliation{
    \institution{Samsung Electronics}
    \country{USA}
}
\author{Sapan Agarwal}
\affiliation{
    \institution{Sandia National Laboratories}
    \country{USA}
}
\author{Saugata Ghose}
\affiliation{
    \institution{University of Illinois Urbana-Champaign}
    \country{USA}
}
\author{Ben Feinberg}
\affiliation{
    \institution{Sandia National Laboratories}
    \country{USA}
}
\renewcommand{\shortauthors}{Wong, et al.}

\makeatletter
\let\@authorsaddresses\@empty
\makeatother

\begin{abstract}
\input{sections/abstract}
\end{abstract}

\maketitle

\input{sections/intro}

\input{sections/background}
\input{sections/framework}
\input{sections/methodology}
\input{sections/database}
\input{sections/graph}

\input{sections/related}
\input{sections/conclusion}

\begin{acks}
\input{sections/sandiastatement}
\end{acks}

\bibliographystyle{ACM-Reference-Format}
\bibliography{refs}

\end{document}

%% file: sections/abstract.tex
As the amount of data produced in society continues to grow at an exponential rate, modern applications are incurring \revSG{significant performance and energy penalties due to high} data movement between the CPU and \revSG{memory/storage.}
\revSG{While processing in main memory can alleviate these penalties,
it is becoming increasingly difficult to keep large datasets entirely in main memory.}
\tRW{\tSG{This has led to a recent push for \emph{in-storage computation}, where processing is performed inside the storage device.}}

\tRW{We propose TCAM-SSD, a new framework for search-based computation inside the NAND flash \tSG{memory} arrays of a conventional \tSG{solid-state drive (SSD),}  \revSG{\tSG{which requires} light\-weight modifications to only the array periphery and firmware.}
TCAM-SSD introduces a \emph{search manager} and \emph{link table}, which can logically partition the \tSG{NAND flash memory's contents} into search-enabled \tSG{regions} and standard storage \tSG{regions}.
Together, these light firmware changes enable TCAM-SSD to seamlessly handle block \tSG{I/O} operations, in addition to \emph{new} search operations, thereby reducing end-to-end execution time and total data movement.}
\tRW{We provide \tSG{an} NVMe-compatible interface \tSG{that provides programmers with the ability to dynamically allocate data on and make use of TCAM-SSD,} allowing the system to be leveraged by a wide variety of applications.}
\tRW{We evaluate three \tSG{example} use cases of TCAM-SSD to demonstrate its benefits.}
For transactional databases, TCAM-SSD can mitigate the performance penalties \tRW{for applications with large datasets, achieving} a \urevRW{60.9}\% speedup over a conventional system \tRW{that \tSG{retrieves data from the SSD and computes} using the CPU}.
For database analytics, TCAM-SSD \tSG{provides} an average speedup of \urevRW{17.7}$\times$ over a conventional system for a collection of analytical queries.
For graph analytics, we combine TCAM-SSD's associative \tRW{search} with a sparse data structure, speeding up graph computing for larger-than-memory datasets by \tRW{14.5}\%.

%% file: sections/intro.tex
\section{Introduction}
\label{sec:intro}

Over the past decade, the amount of data generated and consumed by modern applications has grown exponentially~\cite{reinsel.whitepaper2018}.
\revSG{As one example, the number of photos shared per minute on the Instagram social media application has increased from 3,500 in 2013 to approximately 66,000 in 2022~\cite{Xu2016,Domo2022}.}
This growth has been observed in a wide variety of application domains, including machine learning~\cite{Chowdhery22}, social media~\cite{Forbes, domo}, and business transactions~\cite{Im2018}, \tRW{with the average person producing \SI{102}{\mega\byte} of data per minute~\cite{Domo2023}.}
As the quantity of data grows, increased pressure is placed on existing memory and storage subsystems, as frequent data movement is needed between these subsystems, the on-chip caches, and the CPU.
Unfortunately, this data movement has become a major bottleneck in modern systems, as it consumes large amounts of energy and results in significant performance penalties~\cite{Oliveira2021, Nair2015, Ghose2019, Dally15}. 

To alleviate the overheads of data movement, architects and system designers have been exploring \emph{processing-in-memory} (PIM), a broad field of techniques that can perform computation close to or inside memory and storage devices.
While early proposals for PIM date back to the 1970s, a recent resurgence of PIM has led to significant innovation in the last decade, with solutions proposed across a diverse range of memory technologies (e.g., SRAM~\cite{Aga2017, Eckert2018, Fujiki2019, Jeloka2016}, DRAM~\cite{Gao2021,Hajinazar2021, Seshadri2017, Li2017}, NAND flash \tSG{memory}~\cite{Gao2021, Park2022, Shim2022}, emerging memory technologies~\cite{Angizi2018a, Angizi2018b, Angizi2019, Ankit2019, Chou2019, Gaillardon2016, Gupta2018, Hamdioui2017, Imani2019, Shafiee2016}). 
DRAM-based \tRW{processing-in-memory} has formed the basis of several commercial products available today~\cite{he.isca2020, lee.isca2021, devaux.hotchips2019, upmem.website, samsung.hbmpim.website}.
However, \tRW{\tSG{efforts to increase DRAM capacity continue} to suffer} 
\revSG{from scaling issues that have plagued DRAM manufacturers} over the last 20 years~\cite{mandelman.ibmjrd02, kang.memoryforum14, mutlu.imw13}, and recent reports speculate that these scaling issues will worsen in the coming decade~\cite{Im2020, Litho2020}.
\tRW{These challenges will make it increasingly difficult for main memory capacity to keep up with ever-increasing \tRW{dataset} sizes.}

NAND-flash-based solid-state drives (SSDs) provide an opportunity to perform PIM while overcoming main memory scalability issues.
Compared to DRAM, NAND flash memory offers significantly higher densities \tRW{that} translate into orders-of-magnitude larger capacity at an approximately \tRW{40$\times$} cheaper cost-per-bit~\cite{Neumann2020}.
However, NAND flash access latencies are significantly higher than their equivalents in DRAM.
This makes it more challenging to take near-memory logic originally proposed to sit close to DRAM and \tRW{implement} this logic close to storage.
Instead, to overcome the higher latencies, NAND-flash-based PIM can take advantage of \emph{processing-using-memory} (PUM; a.k.a.\ {\insitu} computing), where we perform logic directly using memory cells to unlock the potential of million-way parallelism (where, generally, a pair of cells can form a compute unit).

Recent works have explored how to implement PUM in NAND flash memory. %
Parabit~\cite{Gao2021} proposes modifications to the latching circuitry used in NAND flash \tSG{memory} to execute bitwise operations between subsequent page-level operations.
Flash-Cosmos~\cite{Park2022} 
extends Parabit by proposing intra- and inter- block bitwise operations, a common operation for a variety of applications, including databases and web search.
GP3D~\cite{Shim2022} accelerates graph analytics, specifically by incorporating a \tRW{PageRank~\cite{Brin1998}} accelerator that leverages the NAND flash array structure to perform analog matrix--vector multiplication.
\revSG{However, despite significant potential for performance and energy improvements, these existing approaches
(1)~have notable limitations that prevent their widespread use (e.g., \tRW{the need to perform} the same \tRW{operation} across thousands of data elements, modifications within the NAND flash array, domain-specific solutions) and
(2)~do not examine how to integrate their low-level in-flash primitives with the larger system and application (e.g., data organization across arrays and across chips, retrieving operand results from the array).}

\revSG{To tackle these issues, we propose TCAM-SSD, a new framework for efficient in-SSD computing.}
TCAM-SSD builds upon \tRW{in-memory \tSG{search}}~\cite{Tseng2020}, a previously-proposed primitive that treats a NAND flash array as a \emph{bulk ternary content-addressable memory} (TCAM).
\tSG{Conceptually, a TCAM searches across multiple data entries in parallel, by driving the search string value (i.e., the search \emph{key}) on wires connected across all of the entries.}
While \tSG{the IMS} work shows how to perform parallel TCAM lookups \tSG{using a NAND flash} array, it does not discuss how to integrate this primitive to perform useful computation for applications, or how the SSD should manage the data and computation \tSG{alongside I/O requests}.
\tRW{
TCAM-SSD provides mechanisms \tRW{that} enable the application to \tSG{(1)~}effectively retrieve data through searchable data elements, \tSG{and (2)~}execute compute operations in the forms of queries.
The proposed framework can be further extended to support in-SSD associative computing, enabling a \tSG{wide range} of applications \tSG{such as} databases~\cite{Caminal2022}, data mining~\cite{Agrawal1994}, network routing~\cite{Pei1991}, \tSG{and} image processing~\cite{Meribout2000}.
}

\tRW{Our goal with TCAM-SSD is to build a full in-SSD computing framework that allows applications \tSG{to} manage data and efficiently make use of IMS.}
IMS implements the TCAM-based search primitive by modifying the voltages applied to each wordline (i.e., row) in the array, \tRW{and by storing data along the bitlines of the NAND flash string.}
\tRW{TCAM-SSD makes four key changes to a conventional SSD to enable applications to effectively utilize IMS.}
First, TCAM-SSD seamlessly enables \tRW{\tSG{searchable} fields within a data region to be stored in \tSG{the SSD using} the column-oriented format required by IMS, while maintaining \tSG{a complete and coherent copy of the} data region in \tSG{the SSD using} the conventional row-oriented format.
This allows TCAM-SSD to enable efficient bulk searches across the data region without requiring the \tSG{programmer} to explicitly manage data transposition.}
Second, TCAM-SSD incorporates an efficient firmware metadata structure for searchable data regions and their results.
\tRW{Third, TCAM-SSD exposes an NVMe 2.0~\cite{NVMeBase2} compliant command interface that applications and/or system drivers can use to perform computing tasks parallelized across many NAND flash arrays.
Together, these changes enable the ability to coherently modify searchable data while maintaining the ability to perform standard SSD operations (e.g., read/write).
}
Fourth, TCAM-SSD introduces multiple optimizations that reduce unnecessary data movement between the NAND flash arrays and the SSD firmware processor.
\tSG{TCAM-SSD requires no modifications to the} NAND flash arrays, \tSG{and requires} only lightweight changes to the array peripheral circuitry and the SSD firmware.

\revSG{While TCAM-SSD can provide performance and data movement improvements across many \tSG{application} domains,
we focus on three examples to demonstrate data structures and semantics that can make use of associative \tRW{search}.}
We evaluate the benefits of TCAM-SSD over a system with a conventional SSD \revSG{for these use cases.}
For transactional databases, TCAM-SSD's associative search can mitigate the performance penalties of accessing the disk, with a \urevRW{60.9}\% speedup.
For database analytics, TCAM-SSD \tSG{provides} an average speedup of \urevRW{17.7}$\times$ for a collection of analytical queries.
For graph analytics, TCAM-SSD combines associative \tRW{search} with a sparse data structure, providing speedups for larger-than-memory datasets of \tRW{14.5}\%.

\revSG{We make the following contributions in this work:
\begin{itemize}
    \setlength\itemsep{0em}
    \setlength{\parskip}{0pt}
    \item \tRW{We introduce the first full framework for in-SSD computing using IMS.}
    \item We modify the SSD firmware to perform associative \tRW{search operations} alongside conventional I/O requests.
    \item We provide an NVMe-compatible interface for applications to perform bulk parallel associative \tRW{search}.
    \item We explore \sg{three use cases to show data structures and application semantics for in-SSD computing}.
\end{itemize}
}

%% file: sections/background.tex
\section{Background}
\label{sec:bkgd}
\label{sec:background}

\subsection{NAND Flash SSD Organization}
\label{sec:bkgd:flash}
\label{ssec:bg-NAND-flash}

Modern \tSG{NAND-flash-memory-based} solid-state drives (SSDs) are typically divided into a front end and a back end.
The front end handles 
(1)~interfacing to the host;
(2)~managing the \emph{flash translation layer} (FTL), the firmware layer responsible for \tSG{mapping the host's logical block address for a piece of data to an SSD-internal physical address}; and 
(3)~dispatching I/O requests to the back end memory subsystem, which contains the NAND flash memory chips.
Due to the complexity of managing these requests and the FTL, 
the front end typically executes on a dedicated microcontroller and includes dedicated DRAM. 

The back end is divided into multiple \tSG{\emph{channels}}, 
which can execute independent I/O operations in parallel.
Additional parallelism is attained by connecting multiple NAND \tSG{flash memory chip} packages to each channel (\tSG{known as way pipelining} or package interleaving).
Furthermore, within each \tSG{chip} package, different \tSG{\emph{dies}} can support interleaved commands (die interleaving).
Although dies are the smallest superstructure \tSG{the front end can issue an independent command to}, each die may contain multiple \tSG{\emph{planes}}, which may \tSG{operate} in parallel when the die is issued specific multi-plane operations. 
Finally, each plane is composed of multiple \tSG{\emph{blocks}}.

\begin{figure}[ht]
	\centering
	\includegraphics[width=\columnwidth]{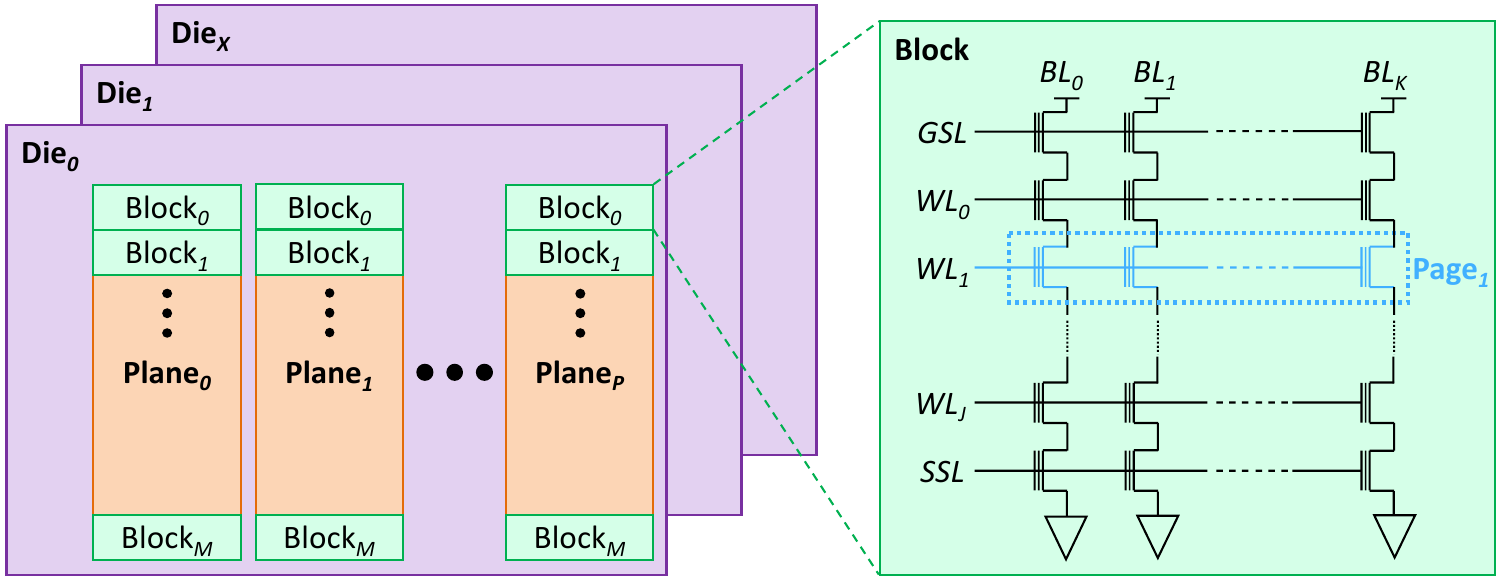}
     \vspace{-10pt}

	\caption{Left) NAND flash organization. Right) NAND flash blocks composed of ground select line (GSL), horizontal wordlines (WL), vertical bitlines (BL), string select line (SSL), and NAND flash cells.}
	\label{fig:bg-NAND-diagram}
\end{figure}

A NAND flash block consists of multiple \tSG{rows of cells}, each of which are formed by connecting \tSG{multiple cells together} via the wordlines as seen in Figure~\ref{fig:bg-NAND-diagram}.
The wordlines \tSG{($WL_j$)} of a block are connected together via shared bitlines \tSG{($BL_k$)}, which connect a vertical column of NAND flash memory cells \tSG{to a shared peripheral circuit that can perform I/O}.
\tSG{Each row contains one or more \emph{pages} of data}, and only a single page per block can be accessed at a time.
Read and write operations are performed at a page granularity, while erase operations are performed on an entire block at once.
To take advantage of the internal parallelism, pages from different channels \tSG{that share the same address offsets} can be connected into a larger virtual structure called a \emph{superpage} \tSG{that spans multiple blocks, with the blocks collectively referred to as a \emph{superblock}}.

A NAND flash cell consists of a transistor with a \tSG{\emph{floating gate}} that can hold charge.
The absence or presence of charge on the floating gate corresponds with the cell \tSG{\emph{state} (i.e., its stored data value)} and sets the threshold voltage (\vth{}) for the cell.
In a \emph{single-level cell} (SLC), the cell can store two states (0 or 1) that correspond to two discrete \emph{windows} of \vth{} levels.
To increase density and reduce the cost per bit, NAND flash manufacturers often store multiple bits in a single cell.
For example, a \emph{multi-level cell} (MLC) can store four states (00, 01, 10, 11) to represent two bits, while \tSG{\emph{triple-level cells} (TLC) and \emph{quad-level cells}} (QLC) can store three and four bits, respectively.

During a read, a voltage is applied to the top of the bitline, \tRW{both ground select and string select lines,} and the state of each cell in a row is read based on whether or not the transistor allows current to flow through it.
To read stored data from \tSG{an SLC NAND flash memory block}, a read reference voltage \vread{}, which is between the two \vth{} levels, is applied to the wordline being read.
If the cell is set to high \vth{}, \vread{} is insufficient to turn on the transistor, resulting in a logical 0 being read out. 
However, if the cell is set to low \vth{}, \vread{} \tRW{turns on the transistor, resulting in a logical 1 being read out}.
To ensure that current flows through the other cells in the same column, all other \tRW{wordlines} are driven to \vpass{}, a voltage sufficient to turn on transistors regardless of the stored state (i.e., \vpass{} $>$ high \vth{}).
For an $n$-bit cell, we need to apply \tRW{up to $n$} read reference voltages for each read, which significantly increases the read latency.
To mitigate this increase, many systems use an SLC cache (i.e., a region of cells that hold only a single bit) to improve performance~\cite{Yang2016}.

To perform a write, \vpass{} is applied to all the wordlines except for the page to be written, which is driven to a programming voltage \vprog{}.
\vprog{} is much higher than \vread{} \tRW{and} \vpass{}, \tRW{which results in electrons being injected into the floating gate}. 
Once electrons have been injected,
it is not possible to remove them until an erase operation is performed.
Therefore, prior to storing data, the cell must be erased by applying a large negative voltage (\verase{}), which 
causes \vth{} of the cell to be set low. 
Multi-bit cells use more complex (and slower) methods to perform programming, such as two-step programming~\cite{Park2008, Cai2014} or foggy--fine programming~\cite{Cai2017, Cai2015b}.

\subsection{Associative Memories}
\label{sec:bkgd:associative}

In many systems, \tRW{searching for specific data elements} can quickly become an expensive operation even with additional metadata (e.g., indexes, hash functions).
Although tree structures provide significant improvement over linearly scanning, the structure still incurs a logarithmic search time. 
Hashing, with constant search time, can lead to collisions that hinder performance.
\tRW{To improve the performance of data retrieval, systems can employ dedicated hardware to provide constant-time searching without aliasing.
A \textit{content-addressable memory} (CAM)~\cite{Batcher1974, Kohonen1980, Wade1987} contains an array of data elements that can be indexed directly by data value (as opposed to the address-based indexing employed by caches).}
\tRW{In general, \tSG{a CAM executes} a lookup \tSG{(i.e., a \emph{query})} in the form of a parallel search across all data elements in an array by broadcasting \tSG{the desired content value (i.e., a \emph{key})}. 
\tSG{The CAM lookup returns a list (often in the form of a bitvector) that indicates which of its entries, if any, contain the key.}}
CAMs are well suited for applications \tRW{in which the location and presence of the data is unknown, e.g., translation lookaside buffers (TLBs), associative caches, database operations~\cite{Sun2017}.}

\tRW{A \emph{\tSG{ternary} content-addressable memory} (TCAM) extends the concept of a CAM by introducing a wild card bit (i.e., a don't care bit; represented as X).
In a TCAM, a bitline whose query value is set to X will match \textit{either} a 0 or a 1 for that bit.}
For example, a single \tSG{TCAM lookup} for 01X0 \tSG{retrieves a list of matches} corresponding to 0100, 0110, and 01X0, \tSG{while a non-ternary CAM would have to perform two lookups} (\tSG{a search for} 0100, \tSG{followed by a} search \tSG{for} 0110).
Due to their additional capabilities, TCAMs have been used in packet classification~\cite{Ravikumar2004}, network intrusion detection~\cite{Graves2019}, and information retrieval~\cite{Li2014}.

\tSG{A TCAM is} typically implemented using SRAM-like structures~\cite{Pagiamtzis2006, Jeloka2016}.
Unfortunately, this comes with significant density and power costs~\cite{Goel2010, Arsovski2003}. 
For every additional \revSG{data element (e.g., word)} in a TCAM, both the static power and the dynamic power per match increase (because a CAM must look up \tSG{\emph{all} of its contained} data elements for every search).
Due to the overheads associated with SRAM-based TCAM, alternative TCAM devices have been proposed using emerging \emph{non-volatile memories} (NVMs)~\cite{Matsunaga2009, Derharcobian2010, Eshraghian2010, Rajendran2011, Xu2010, Alibart2011, Matsunaga2011, Yin2019, Zha2020, Narla2023}. 
NVM-based TCAMs are significantly more area- and power-efficient~\cite{Graves2020}.
\tRW{In this work, we focus on implementing TCAM using NAND flash memory, where we can attain a high storage density using more mature technology than emerging NVMs, with scalability and static power benefits over SRAM-based TCAM.}

%% file: sections/framework.tex
\section{The TCAM-SSD Framework}
\label{sec:framework}

We introduce a framework for performing associative \tRW{search} inside NAND-flash-\tSG{memory-}based solid-state drives (SSDs), which we call TCAM-SSD. 
By enabling \tRW{in-SSD} associative \tRW{search}, TCAM-SSD can quickly identify relevant pieces of data without having to send the entire dataset to the CPU, significantly reducing the I/O traffic required for modern applications to manage and process very large datasets. 
\revSG{Unlike prior works, TCAM-SSD enables in-NAND-flash computation without making any modifications inside the NAND flash arrays \revSG{within a NAND flash block}, and exposes an interface that is compatible with popular SSD protocols (and uses existing protocol hardware).
In hardware, TCAM-SSD requires only minimal modifications to the array periphery to support the in-array search.} We modify the embedded firmware to logically partition the SSD into \emph{data regions}, which store file blocks 
in a conventional manner, and 
\emph{search regions}, where we use an efficient transposed data layout to enable rapid, highly-parallel associative search \revSG{across data elements}.

\subsection{High-Level Overview}
\label{sec:framework:overview}

\begin{figure}[h!]
    \centering
    \includegraphics[width=.6\columnwidth]{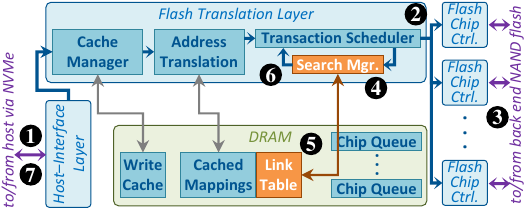}
    \caption{TCAM-SSD front end (modules \tSG{introduced for TCAM-SSD are shown} in orange).}
    \label{fig:framework-front}
\end{figure}

Figure~\ref{fig:framework-front} shows the front-end interface for TCAM-SSD.
\sg{TCAM-SSD aims to eliminate two types of data movement required by
conventional drive reads:
CPU--FE (front end) and FE--BE (back end).}
Applications interact with TCAM-SSD through \revSG{drive-level} commands that we introduce as extensions to the standard NVMe protocol.
An application uses one of these commands to allocate a new search region.
Our modified FTL performs block-level allocation for the search region, and \revSG{data elements are} written to the NAND flash chips in a vertical manner (i.e., the bits of a word are written to the same bitline of a NAND flash block, with the bits distributed across different wordlines).
TCAM-SSD provides additional commands that can update \revSG{a data element}, and can append new data to a search region. 
These regions, and the keys that they contain, are efficiently tracked in the firmware.

To perform a \emph{ternary} search (i.e., a search that either looks for a matching value for each bit or treats a bit as a don't care) over one or more search regions, the application issues a search command over NVMe to the firmware (\ballnumber{1} in Figure~\ref{fig:framework-front}). 
The firmware schedules search commands for the NAND flash chips in the back end (\ballnumber{2}). The command uses per-wordline read reference voltages to represent each bit of the search \revSG{key} (or whether a bit is a don't care), and passes these voltages to the NAND flash blocks that contain the search regions requested by the command (\ballnumber{3}).
\revSG{The NAND flash blocks, using modified peripheral circuitry to support per-wordline voltages, then issues a chip-level \emph{SRCH} command that can concurrently search through thousands of data elements in the block at once.}
The output of each bitline indicates whether the word stored along the bitline is a match.
\revSG{Combined with block-level parallelism, a single  \emph{SRCH} command} can search over tens of thousands of \revSG{data elements} simultaneously.
The list of matches is returned to a search manager that we add to the firmware (\ballnumber{4}), which uses a metadata table to decode where the specific \revSG{data lies} (\ballnumber{5}).
The firmware then schedules and issues read requests for only the matching data (\ballnumber{6}), and \revSG{the matching data is} returned to the host (\ballnumber{7}).

\subsection{Implementing the Search Primitive}
\label{sec:framework:primitive}
\label{ssec:tcam-array}

We \revSG{implement a primitive based on IMS~\cite{Tseng2020}} for \emph{ternary search} (i.e., for each bit, match a 0 or 1, or ignore that bit)
within a conventional NAND flash \revSG{block}.
We do this by modifying the layout and programming pattern of the cells in a NAND flash block,
and by changing the \vread{} and \vpass{} voltages that are applied to the \revSG{block's wordlines}.
\emph{Note that our approach requires zero changes to the layout or design of \revSG{the cell array within the block}.}
Even with our changes,
\revSG{a block} can still perform conventional read, program, and erase operations,
allowing \revSG{it} to continue functioning as storage.

\begin{figure}[h!]
  \centering
  \includegraphics[width=0.6\columnwidth]{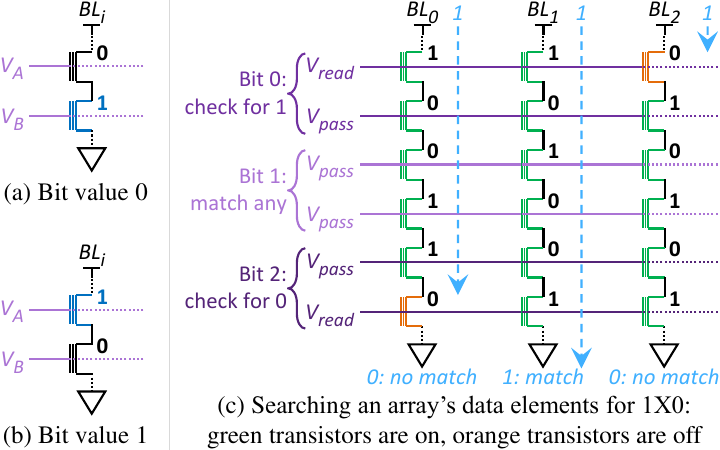}
  \caption{\revSG{\sg{Associative} search in a NAND flash array.}}
  \label{fig:tcam-data-values}
  \label{tab:ki-TCAM-cell}
\end{figure}

\paragraph{Storing a Bit}
To enable ternary search, we logically (but not physically) combine two NAND flash cells,
which share the same bitline and sit in adjacent rows to one another, to represent a single bit of data.
\revSG{We can use these two cells
to store a \revSG{bit} value 0 (Figure~\ref{fig:tcam-data-values}a), a \revSG{bit} value 1 (Figure~\ref{fig:tcam-data-values}b),} or a \revSG{bit} value X (representing a bit that will match
either a 0 or a 1; i.e., a bit that can be ignored).
For each NAND flash cell, we use the same \vth{} states 
\revSG{as conventional} SLC cells.\footnote{\sg{While TCAM-SSD can make use of multi-level cells, we use SLC, without loss of generality, to simplify our descriptions.}}

\paragraph{Mapping Data in a Block} \label{p:framework-mapping}
\revSG{Unlike conventional SSDs, which store a bit of data in a single cell
and store all bits of a data element across cells that share a wordline within a NAND flash block,
TCAM-SSD transposes how data is stored for blocks within a search region.\footnote{\revSG{Blocks in data regions \sg{still} use the conventional data mapping.}}
All of the bits of a data element are stored along the same bitline, 
with each bit stored across two adjacent cells.}
For a typical NAND flash block where the block spans 128--512 rows~\cite{Cai2017},
this allows us to store \revSG{data elements} as large as \revSG{63--255 bits, with the last bit used to represent whether the element is valid}.\footnote{Note that we can store and search for shorter \revSG{data elements}, by setting
the search \revSG{key} bits to X for the bits that do not belong to the \revSG{data elements}.}

To reduce the complexity of programming and interacting with our transposed representation,
and to avoid introducing cell-to-cell program interference,
we allocate an entire block of NAND flash memory at once for those
blocks that we will perform search on.
Despite \revSG{performing block-level allocation,}
TCAM-SSD uses existing page-level programming operations
to write \revSG{data to the block, handling} transposition transparently to the programmer 
(\revSG{see} Section~\ref{sec:framework:ftl}).

\paragraph{Checking \revSG{a NAND Flash Block for Matches}}
Our \revSG{transposed organization} allows us \revSG{to search an entire block at once,}
by using read reference voltages to represent the \revSG{bits of the search key}.
\revSG{Figure~\ref{fig:tcam-data-values}c shows an example for a $6 \times 3$ NAND flash block.
Each column represents a different data element, with pairs of rows
representing one bit of the element.
Across Bit~0 (i.e., the first two rows) of each data element,
we search for a bit value 1 by applying \vread{} to the first row
and \vpass{} to the second row.
Bit~1 shows how to ignore a bit (i.e., match either a 0 or a 1),
and Bit~2 shows how to search for a bit value 0.
For a given search key, TCAM-SSD chooses and applies
the correct wordline voltages for each bit in the key,
using a new chip command that we introduce called \emph{SRCH}.
\emph{SRCH} contains one bit per wordline to indicate the search key
(i.e., whether to apply \vread{} or \vpass{}),
and we replace the block's voltage selection decoder with
per-wordline 2:1 muxes, 
similar to proposals from prior work~\cite{Cai2012, Cai2013, Cai2015a, Cai2017b}.}

\revSG{When \emph{SRCH} is performed on a block, a bit value 1 is
applied at the top of each bitline, identical to how reads are
performed.
If all bits of a data element match the search key,
all transistors along the bitline turn on, and
the 1 propagates to the bottom of the bitline.
If \emph{any} bit does not match, one of that bit's transistors
will be off (e.g., orange transistors in Figure~\ref{fig:tcam-data-values}c), 
and the 1 \emph{stops} propagating.
\emph{SRCH} searches all data elements in a block \emph{simultaneously}, and 
returns a \emph{match vector} (i.e., the bitline outputs) as a single value.
Note that the last cell in each bitline stores
a valid bit (0 if valid, 1 if invalid; not shown in the figure), and \vread{} is applied to the
this wordline, preventing invalid elements from matching.}

\subsection{Managing Searchable Data in Firmware}
\label{sec:framework:ftl}
\label{sec:framework:firmware}
\label{ssec:ftl}

To maintain the ability to perform baseline SSD functionality and as well as our new associative search functionality, TCAM-SSD splits the SSD into two types of regions, \emph{data} and \emph{search}.
Data regions behave exactly the same as data currently does in a conventional SSD, relying on conventional reads, programming, and erase.
Search regions can be allocated only through a new command that we introduce (Section \ref{ssec:tcam-alloc}),
where the size of the region is based on the number of \revSG{data elements and the element length.}

Note that we do not limit the length of a \revSG{data element} to the number of NAND flash cells along a bitline in a block ($\frac{\text{wordlines per block}}{2}$, which we define as the \emph{native \revSG{element} size}).
In cases where we need longer \revSG{data elements}, a search region can be extended across multiple blocks,
\revSG{with each data element split across the blocks}.
\revSG{While this may require additional search operations,}
the resulting match vectors from each block can be ANDed together to form a final match vector prior to decode and accessing the data region.
Alternatively, for large tables, the number of keys in a search region may exceed the number of columns in a single block (page size) and the match vectors must be concatenated together.

\paragraph{\revSG{Linking Search Regions to Data Regions}}
Each search region is connected to a data region, and the mapping between the regions is maintained in a software-controlled \emph{link table}.
For each data element in the search region, the data region contains a corresponding \emph{data entry}.
Both data elements and data entries have a fixed length, allowing the link table to store only a single base physical address per block in the \rw{data} region (along with a pointer to a firmware buffer for updated values; see Section~\ref{sssec:tcam-update}).
The TCAM-SSD firmware can add an offset to the base address to look up specific data entries corresponding to matching data elements.

\revSG{The contents of the linked data entry are application dependent.
For example, if an application directly wants the value of the matching data element, its corresponding data entry contains a
non-transposed replica of the element value.\footnote{By replicating the data in a conventional wordline-oriented data region, we can \revSG{read an $n$-bit data element's value out in a single cycle, instead of performing $n$~reads (one for each bit) from the transposed search region.}}
If we want to implement a key--value store (KVS), which stores tuples of keys and values,
each data element in the search region would correspond to the key,
and its corresponding entry in the data region would hold the value (and, if desired, a non-transposed copy of the key).}

\paragraph{Support for Block-Level Allocation} 
\label{p:ftl}
TCAM-SSD requires modifications to the flash translation layer (FTL) to implement search.
Most importantly, rather than using page-level allocation, 
\revSG{search regions (but not data regions)} use block-level allocation since pages within a search block must be allocated contiguously.
 Notably, this block-level allocation requirement only applies to search\:regions: data regions can continue to use the underlying FTL implementation.
In a conventional SSD, superblocks are formed from a collection of blocks usually from different chips with the same block offset.
Accordingly, prior work has proposed superblock FTL designs~\cite{Jung2010}.
TCAM-SSD is amenable to this type of allocation scheme, as it enables the system to search over an entire superblock in parallel.

\subsection{TCAM-SSD Command Interface}
\label{sec:framework:cmds}
\label{ssec:tcam-command-set}
To interface with TCAM-SSD, we propose a TCAM-specific NVMe 2.0 compliant command set specification~\cite{NVMeCMD2}.
The proposed commands are similar to the KVS command set ratified as part of NVMe 2.0~\cite{NVMeKVS}, and take advantage of the ability to add vendor-specific functionality to the interface.
The commands described below are sufficient to implement a basic set of associative computing functionality; however, more advanced commands may be needed to support a wider range of functionality in the future
(e.g., updating data \revSG{elements in place}).

\paragraph{Allocate / Deallocate / Append}
{\label{ssec:tcam-alloc}}
Since \revSG{search} regions are managed separately from data regions, they must be specially allocated and managed.
The \emph{Allocate} command creates a \revSG{search} region based on the data \revSG{element} size, and a linked data \revSG{region based on the data entry size,} as discussed in Section~\ref{sec:framework:ftl}.
The command can optionally provide data for the search region by providing a pointer to the host memory \revSG{that stores elements and entries. The \emph{Deallocate} command frees a search region by marking all blocks for erase.}

The \emph{Append} command is used to add \revSG{elements of the same size to an existing} search region, \revSG{along with their corresponding data entries to the data region.
The firmware stores the new data element (along with its corresponding data entry) in a software buffer.
Once there are enough elements in the software buffer to fill an entire block, one new block each is appended to the search region and the data region,
and the elements and their data entries are transferred out of the buffer and written to the drive.
The firmware appends the link table with the new mapping.}

Importantly, allocate and append require that the order \revSG{of data elements is the same as the order of data entries} in the linked data region.
This requirement must be managed by the host application.

\paragraph{Simple Search / Search / Search Continue}
\label{sssec:tcam-search}
Once \revSG{data elements and corresponding data entries are written,}
the host can issue search commands to the SSD.
For simplicity, we describe the \emph{Simple Search} command, which contains \revSG{a fixed-length search key (up to 127~bits), the address of the search region, and a pointer to a host buffer where return data can be stored}.
Alternatively, if the native \revSG{element} size exceeds 127~bits, the host may issue a \emph{Search} command, which uses a data pointer to communicate the search \revSG{key} to the SSD.
Both \tSG{the} \tRW{\emph{Search}} and \tRW{\emph{Simple Search}} commands can also communicate additional operations to complete with the resultant search data, such as logical AND and OR reductions between shorter keys.
\revSG{The firmware issues one or more \tSG{chip-level} \emph{SRCH} commands to the \tSG{selected} chips to perform the bulk parallel search, \tSG{where each command invocation returns} a match vector.
\tSG{The firmware uses each} match vector, along with a base address from the link table, to calculate the addresses for data entries in the data region that correspond to matches.
The firmware then issues read commands to these addresses, and writes the returned data to the host buffer.}

\revSG{Note that the} \tSG{\emph{Search}} command does not know how many tuples will match, and therefore may not allocate sufficient host buffer space to store all of the returned \revSG{data entries}.
To address this, a flag is added to the completion queue entry, notifying the host that the host buffer was inadequately sized.
Upon receiving this signal, the host may issue a \emph{Search Continue} to the same search region address with a new host buffer to complete the 
data transfer issued by a prior search.

\paragraph{Delete}
\label{sssec:tcam-update}
\revSG{\sg{When the \tRW{\emph{Delete}} command is invoked to remove} a data element and its corresponding data entry,
the firmware first \revSG{searches for the data element} in the search region.
The firmware then invalidates all matching data elements, by using normal chip-level page commands to read and update the element valid bits
for each block containing a match.
This invalidation is written \emph{in place}, since it involves only raising \vth{} of one cell per match from a 0 to a 1.}

Updating 
an existing \revSG{data element (or its associated data entry)} involves \revSG{first calling \tRW{\emph{Delete}} to remove the old value, and then calling \tRW{\emph{Append}}}.
\revSG{While such updates are costly, we find that they are infrequent (or non-existent) for many target applications, which tend to use relatively stable datasets.}

\input{sections/software} %
\input{sections/hardware}

\subsection{Comparing In-Storage Compute Techniques}
\label{sec:framework:comparison}
\label{sec:framework:datamovement}
\label{ssec:integrating}
\sg{As discussed in Section~\ref{sec:framework:overview}, TCAM-SSD aims to eliminate unnecessary CPU--FE and FE--BE data movement.
Prior works have proposed two families of in-storage compute techniques that also target these types of data movement.}

\sg{\emph{Computational SSDs} (e.g., \cite{samsung.smartssd.website, samsung.smartssd.gen2.pressrelease}) reduce CPU--FE data movement by introducing dedicated compute logic (e.g., FPGAs, embedded IP cores)
in either the front end of the SSD or just outside of the SSD's host interface.
Because the introduced logic does not have direct access to the back end NAND flash chips,
computational SSDs cannot eliminate FE--BE movement.}

\sg{\emph{In-flash bitwise processing} (IFBP) techniques target both CPU--FE and FE--BE data movement
by performing processing using NAND flash memory cells.}
\rw{Flash-Cosmos~\cite{Park2022} is a state-of-the-art \sg{IFBP} technique that performs \sg{bulk Boolean} operations.
\sg{By focusing on Boolean operations, Flash-Cosmos restricts its opportunities to reduce FE--BE movement compared to TCAM-SSD.}
For example, for a table containing 64-bit user entries and a list of accessed websites, 
\sg{Flash-Cosmos} can determine \sg{the active user count for a given website using PUM, 
but cannot use PUM to answer \emph{which} users are active (instead requiring data to be sent back to the CPU);
TCAM-SSD can perform both using memory.}}

\sg{While there is some overlap between the functionality of computational SSDs, Flash-Cosmos, and TCAM-SSD,
they offer some complementary benefits, and TCAM-SSD can operate in conjunction with both.}
\rw{%
\sg{For example, an application can use TCAM-SSD to execute} a portion of the computation within the \sg{NAND} flash arrays, reducing \sg{CPU--FE and FE--BE} data movement \sg{compared to a CPU-based search}. 
\sg{Computational SSD logic can then post-process the search results returned by TCAM-SSD, further reducing CPU--FE movement.}}
\rw{Similarly, the Flash-Cosmos inter-block `OR' operation can be leveraged by TCAM-SSD to enable search of data elements larger than the native element size.}

\subsection{Targeting Use Cases for TCAM-SSD}
\label{sec:framework:apps}
\label{sec:framework:usecases}

TCAM-SSD can be broadly applied to a variety of application domains that benefit from associative \tRW{search}.
Examples include association rule mining~\cite{Agrawal1994}, image processing~\cite{Meribout2000}, database analytics~\cite{Caminal2022}, hardware reconfiguration~\cite{Zha2018, Zha2020}, and text processing~\cite{Imani2016}.
Associative memories have also found uses in packet classification~\cite{Spitznagel2003, Lakshminarayanan2005}, IP routing~\cite{Zane2003}, and network intrusion detection systems~\cite{Chang2008}.
In this work, we demonstrate how to apply and optimize TCAM-SSD for three example use cases:
(1)~mitigating the disk access overheads of online transaction processing in databases (Section~\ref{sec:db:oltp}),
(2)~reducing the data transferred between storage and the CPU for database analytics (Section~\ref{sec:db:olap}), and
(3)~improving the processing efficiency of \ben{large} graphs during graph analytics (Section~\ref{sec:graph}).

%% file: sections/software.tex
\subsection{Software Support}
\label{sec:framework:software}

\urevSG{We wrap the TCAM-SSD NVMe commands from Section~\ref{sec:framework:cmds} into a programmer-friendly API.
This API can be used in two modes:
(1)~NVMe Mode, where data is returned to the host CPU for computing; and
(2)~Associative Update Mode, where data is not transferred over NVMe, and bulk updates are performed directly in the SSD on matching data.}

\paragraph{NVMe Mode}
Listing~\ref{code:software:NVMe_mode} shows an example application
\urevSG{where we search a dataset for all salary records of employees named ``Bob''.}
Once the data entry is brought to the host, the programmer is able \urevSG{to modify the matching records (e.g., give all Bobs a raise), and send the updates back to the SSD.}

\begin{minipage}[t]{.45\textwidth}
\begin{lstlisting}[
    label={code:software:NVMe_mode},
    language=C++,
    basicstyle=\ttfamily\footnotesize,
    keywordstyle=\color{blue},
    stringstyle=\color{red},
    commentstyle=\color{JungleGreen},
    breakatwhitespace=false,         
    breaklines=true,                 
    keepspaces=true,                 
    numbersep=5pt,                  
    showspaces=false,                
    showstringspaces=false,
    showtabs=false,                  
    tabsize=2,
    caption=NVMe Mode,
]
Array employees, firstName;

// create Search Region
void *sr =
  AllocSearchable(firstName, employees);

// retrieve data for any Bob from SSD
void *dataEntry =
  SearchSearchable(sr, "Bob");

// update salaries for all Bobs in host
UpdateSalary(dataEntry, newSalary);

// send modified salaries back to SSD
UpdateSearchVal(sr, "Bob", dataEntry);
\end{lstlisting}
\end{minipage}\hfill
\begin{minipage}[t]{.45\textwidth}
\begin{lstlisting}[
    label={code:software:AC_mode},
    language=C++,
    basicstyle=\ttfamily\footnotesize,
    keywordstyle=\color{blue},
    stringstyle=\color{red},
    commentstyle=\color{JungleGreen},
    breakatwhitespace=false,         
    breaklines=true,                 
    keepspaces=true,               
    numbersep=5pt,                  
    showspaces=false,                
    showstringspaces=false,
    showtabs=false,                  
    tabsize=2,
    caption=Associative Computing Mode,
]
// variables are pre-allocated 

// matches for Bob are moved to SSD DRAM
void *bA =
  SearchSearchable(sr, "Bob", capp=true);

// sets desired operation for each match to add 1
void op1 = +1;

// perform modification to each match in SSD DRAM
UpdateSearchVal(sr, bA, op1, capp=true);

DeallocSearchable(sr, capp=true);
\end{lstlisting}
\vspace{10pt}
\end{minipage}

\paragraph{Associative Update Mode}
Listing~\ref{code:software:AC_mode} shows the same example using associative \tSG{update} mode. 
In this mode, the search operation brings the matching entries for ``Bob'' into a DRAM buffer inside the SSD. 
Using the associative \tSG{update} command, the programmer can send an operation (e.g., addition) and immediate to the SSD, which will be bulk update all matching records. 
Notably, this mode does not require moving the records between SSD and host.

%% file: sections/hardware.tex
\subsection{Hardware Optimizations}
\label{sec:framework:hw}
\label{sec:hardware}

\sg{We propose \rw{four} hardware optimizations to improve the efficiency of TCAM-SSD.}

\subsubsection{{Enhancing Reliability}}
\label{sec:framework:hw:slc}

\sg{There are several sources of errors that can induce bit flips
in data read out from NAND flash chips~\cite{Cai2017, Shin2012, Matsui2017, Huang2014}.
While our in-SSD operations intentionally avoid sending data to the
firmware microcontroller to minimize \sg{FE--BE} data movement, this
prevents the operations from using the firmware-based
error correction techniques to mitigate these read errors,
which can introduce false positives and false negatives. %
TCAM-SSD reduces the impact of these read errors in two ways.}

\sg{First, TCAM-SSD operations are expected to induce significantly fewer
read disturb errors (and, thus, false negatives) than read operations.
A key factor on the magnitude of read disturb errors is the
value applied to the wordline.
During a read operation\rw{, \sg{the firmware applies \vread{}}} to one row in 
an $n$-row NAND flash block \rw{and}
\vpass{} to the other (i.e., $n-1$) rows.
During our search operations, \sg{the firmware applies \vpass{}} to only
$n/2$ rows, with \vread{} applied to the other $n/2$ rows.\footnote{\tSG{A wild card search (i.e., setting one bit to don't care, or X) replaces one \vread{} with an additional \vpass{}}.}
\vread{} is typically 2--3$\times$ smaller than \vpass{}~\cite{Cai2017},
and prior work has shown that this lower wordline voltage leads to
an exponential drop in read disturb induced errors~\cite{Cai2015b}
(e.g., a 1300$\times$ drop in errors from just a 6\% drop in voltage).}

Second, we employ the enhanced SLC-mode programming (ESP) technique proposed in Flash-Cosmos~\cite{Park2022} with the goal of eliminating retention errors (and, thus, false positives), for the search region.
ESP treats NAND flash cells as SLC, and increases the \tRW{$V_{th}$ margin} between bit value 0 and bit value 1, which reduces the probability that the voltage of a cell programmed to bit value 0 will drop below \vread{} and be incorrectly read as bit value 1.
\tRW{Notably, prior work has reported that the error rate increases with the number of bits per cell; specifically, Huang et al~\cite{Huang2014} observe that the MLC error rate is 2 orders of magnitude greater than SLC.
Therefore, ESP can substantially improve reliability.}
While this decreases the bit density of the SSD, we do this only on the search region cells, while data regions (non-TCAM-SSD cells) continue to use multiple levels per cell.\footnote{\tRW{Non-search region MLC/TLC cells continue to use conventional error-correcting codes (ECC)}.}

\tRW{
SSDs usually optimized for high-capacity, relying on multi-level cells to increase the bit-density of the drive.
TCAM-SSD's reliance on SLC cells can therefore adversely affect total drive capacity;
however, only search regions require SLC cells and the linked data region can continue to use MLC/TLC.
Thus, the impact on capacity comes \textit{solely} from the new \textit{search regions}, which we quantify in our evaluation.
Additionally, prior work~\cite{Jimenez2013} has explored using some arrays are used as an SLC cache---reducing the effective density penalty to only 2$\times$ within the search regions.
We generalize to all SSDs and report the search region overheads in terms of blocks of total SSD.
}

\subsubsection{Supporting Early (Conditional) Termination}
\label{sec:framework:hw:earlyterm}

Once a search has been executed on the NAND flash arrays, the data must be read back by the microcontroller. 
However, this data is a \tRW{match vector}, which must be decoded prior to executing the secondary read. 
Because we are searching for a particular piece of data among a large dataset, we expect the majority of the match vectors to return \tSG{no matches.}
Therefore, storing a match vector of 0s and subsequently decoding 0s to their corresponding value would waste \tSG{the limited in-SSD DRAM capacity}.
Accordingly, we propose a mechanism to support early (conditional) termination.

At each flash channel controller, we add a small circuit to quickly decode the match vector as it is read from the back end.
In particular, if the data burst is all zeroes, we increment a counter and discard the data burst.
If there is a match, then we tag the burst with the value stored in the counter, which is used to later decode the corresponding value.

\subsubsection{Write Inversion}
\label{sec:framework:hw:writeinv}

Modern NAND flash supports inverse reads, in which the data read from the NAND flash device is the inverse value that is stored in the cell~\cite{Park2022}.
While initially designed for reads, this functionality can be used as part of the program--verify operation to accelerate writes~\cite{Lee2002b}.
We can make use of write inversion to reduce the amount of command data transmitted to the NAND flash chips.
If we restrict our primitive from Section~\ref{sec:framework:primitive} to store only \revSG{bit} values 0 and 1 (i.e., we no longer store X values in the SSD,\footnote{In our evaluated use cases, we do not find a need to store X values.} but can still use X as a don't care in a search value), the two wordlines sharing a single data value in TCAM-SSD are the inverse of each other.
Once the program operation for a wordline has been completed using a program--verify operation, a subsequent row address can be supplied without additional write data.
If the program operation is then executed, the inverse data will then be written to the new wordline.
Note, that by using write inversion, we can reduce 
\revSG{data movement between the firmware processor}
and the NAND flash arrays by approximately 2$\times$ \revSG{during programming}.

\subsubsection{Data Result Compaction}
\label{sec:framework:hw:compact}
\label{ssec:tcam-compact}

Associative search operations may return multiple matches.
However, the corresponding \revSG{matches may not be contiguous} in the data region.
Therefore, a search operation with \emph{N} matches would return \emph{N} \revSG{logical blocks (i.e., pages)} of data to the host, resulting in wasteful data movement.
Thus, if the \revSG{maximum data entry size} is less than the size of a host logical block, we compact multiple data \revSG{entries} into a few host blocks, which are returned to the host.
To enable this optimization, it is necessary to know the size of the corresponding data \revSG{entries} in the data region.
This information is provided to TCAM-SSD as part of the Allocate command, and \revSG{is stored in} the link table structure.

%% file: sections/methodology.tex
\section{Experimental Methdology}
\label{sec:meth}
\label{sec:experimental}

To evaluate TCAM-SSD, we develop a set of detailed analytical models to capture the low-level hardware and software modifications to the SSD, as well as all critical aspects of the system\urevRW{, including, NVMe initialization overheads, DRAM access times, and NAND flash access time.
For example, the latency of the \textit{SRCH} operation (Section~\ref{sec:framework:overview}) includes, NVMe (\ballnumber{1} in Figure~\ref{fig:framework-front}),  translation (\ballnumber{2}), block-level search to the search region (\ballnumber{3}), match-vector retrieval and decode (\ballnumber{4} \& \ballnumber{5}), physical access(es) to the data region (\ballnumber{6}), FE--BE data movement, and CPU--FE data movement (\ballnumber{7}).
}
\urevRW{Additionally, we include support channel- and die-way parallelism.}

Table~\ref{tab:exp-parameters} lists the parameters for the evaluated \revRW{3D NAND flash SSD}, \revSG{based on the configuration in Flash-Cosmos~\cite{Park2022}} \urevRW{including the write latency for ESP}.
\urevRW{The NVMe initialization overhead is set to \SI{4}{\micro\second}, based on prior work~\cite{Tavakkol2018, Lawley2014, Liu2004}.}
\tRW{All other parameters are matched to Flash-Cosmos for ease of comparison.}

\begin{table}[h]
\caption{\revRW{3D NAND flash configuration}~\cite{Park2022}.}
\label{tab:exp-parameters}
\centering
\small
\begin{tabularx}{\columnwidth}{|l|X|}
\hline
\textbf{Parallelism} & channel $\times$ die \\
\hline
\revRW{\textbf{Back End}} & channels = 8,
packages/channel = 1, 
dies/package = 8,
planes/die = 2,
blocks/plane = 2,048,
pages/block = 196,
page size = \SI{16}{\kilo\byte} \\
\hline
\textbf{DRAM Access Time} & \tSG{\SI{15}{\nano\second} to retrieve \SI{64}{\byte}} \\
\hline
\textbf{Flash Access Times} & read = \SI{22.5}{\micro\second}, search = \SI{25}{\micro\second}, write (SLC/MLC/TLC) = \SI{200}{\micro\second}/\SI{500}{\micro\second}/\SI{700}{\micro\second} \\
\hline
\textbf{Max Search Size} & \SI{128}{\kilo\nothing} keys $\times$ channels $\times$ dies, \rw{native element size}\ = \SI{97}{\bit}\\
\hline
\end{tabularx}
\end{table}

We conservatively assume that the data in the data region for both the baseline and TCAM-SSD \sg{resides} in SLC portions of the SSD. 
This makes the search and read latencies comparable; in contrast, if the data resides in \sg{an} MLC cell, the read latency would be significantly larger than that of the search operation, as TCAM-SSD uses SLC to enable search.
Our model captures the effect of channel\urevRW{- and die}-level parallelism, allowing multiple in-flight operations.
 across different channels.
The approximate latency for standard NVMe operations (e.g., read, program, erase) are computing by using the latency of NVMe, translating block addresses to physical addresses, physical page access, FE-BE data movement, and CPU-FE data movement.
New TCAM-specific commands incur the overheads of the the respective operations.
\urevRW{
Our analytical model makes two conservative assumptions that adversely affect TCAM-SSD.
First, we assume that the NAND flash access time for search is $\sim$10\% higher than a read operation\footnote{\tSG{The IMS~\cite{Tseng2020} work notes} that for a block with 96 wordlines and \SI{128}{\kilo\nothing} bitlines, a reliable search operation can be executed within the read latency}
and reflects the increased read latency observed in Flash-Cosmos~\cite{Park2022} (3.3\% for intra-block MWS and 3.3\% for \tRW{four-block} inter-block MWS).
The baseline read operations are unaffected.
Second, multi-block SRCH operations block all internal parallelism needed by the SRCH operation, even when the result is a single data entry.
Search regions may span multiple blocks (Section~\ref{ssec:ftl}), requiring a SRCH operation to check all blocks in the region; however, retrieving the associated data entry may only require a single read operation.
Conservatively, we block all occupied channels/dies required by the search operation even when a single match is required.
Baseline reads are unaffected.
}

We evaluate two different classes of applications, databases and graphs.
For databases, we use DBx1000~\cite{Yu2014} to generate traces of TPC-C~\cite{TPC-C} behavior. 
To evaluate TCAM-SSD for database analytics, we examine queries evaluated in prior work~\cite{Gu2016, Woods2014}.
Similarly, we use dbgen~\cite{dbgen} to populate a database with data specified by TPC-H~\cite{TPC-H}.
Finally, to evaluate graph workloads, we extract a vertex traversal of a single-source shortest path (SSSP) algorithm~\cite{Malicevic2017}, and use a collection of synthetic and real-world graphs ranging from geographical data to social networks (see Section~\ref{sec:graph}).

%% file: sections/database.tex
\section{Evaluating TCAM-SSD for Databases}
\label{sec:db}
\label{sec:database}

\sg{Many applications make use of \emph{relational databases}, which store a series of \emph{records} into one or more \emph{tables}.
Each table consists of one or more columns, where each column corresponds to a particular attribute,
and each record (i.e., row) in a table contains a value per column.
For example, a record for a college ID system could contain a student's ID number, first name, and last name,
and the table would hold records for every student (or, depending on the database design,
have per-department tables of students).
The goal of a database is to allow users to quickly retrieve information about the stored records.
There are two common approaches to retrieval.
\emph{Transaction processing} searches particular columns in one or more tables, and returns or updates individual records whose attributes match the search query.
\emph{Analytics processing} scans entire tables and extracts aggregated information about one or more columns across all records.}

\sg{Relational databases are designed to handle large amounts of data.
Scanning through each record in the table to identify matches for a transaction can be time consuming and require a significant amount of I/O.
To reduce these costs, database managers generate \emph{indexes}, which use data structures (e.g., hashes, trees) designed for sub-linear search/lookup times across all of the records, to catalog the contents of one or more table columns~\cite{Lehman1985, Zhang2018}.
A table can have one \emph{primary index}, which contains unique pointers for each record in the table,
and multiple \emph{secondary indexes}, whose pointers may correspond to multiple records (and thus require additional processing to retrieve matching records).
Note that the exact choice and structure of indexes depend heavily on the workload, and on the specific design of the database.}

\sg{Ideally, all of the contents of a database (i.e., all indexes and all tables) would fit in main memory,
to reduce the latencies of index traversal and data lookup~\cite{Stonebraker2013, Kallman2008}.
However, as datasets continue to grow, it is becoming increasingly difficult to keep the tables
resident in memory, and for large enough databases, main memory will not even be able to hold all of the indexes.
In a conventional system, this would result in a significant latency penalty for both transactions and
analytical queries, as multiple SSD read operations would be required.
We propose to use TCAM-SSD to significantly reduce the latencies involved with large-footprint databases.}

\subsection{Online Transaction Processing Workloads}
\label{sec:db:oltp}
\label{ssec:db-oltp}

\sg{An \emph{online transaction processing} (OLTP) workload provides data in response to an end user executing a series of transactions~\cite{Kemper2011, Yu2014},
where each transaction consists of a series of small, simple queries.
\rw{\sg{OLTP workloads are often generated by user interaction, and thus require high throughput \emph{and}} low latency,}
\sg{so database designers} attempt to fit the entire database in memory~\cite{Stonebraker2013, Kallman2008}.
Unfortunately, this is difficult to do for many larger databases, and
while several optimizations attempt to maximize main memory usage in such scenarios (e.g., \cite{Diaconu2013, Eldawy2014, DeBrabant2013}),
metadata pressure and growing dataset sizes undermine these optimizations and often induce a high amount of SSD I/O operations~\cite{Leis2018, Stoica2013, Zhang2016}.}

\begin{figure}[!h]
	\centering
	\includegraphics[width=.9\columnwidth]{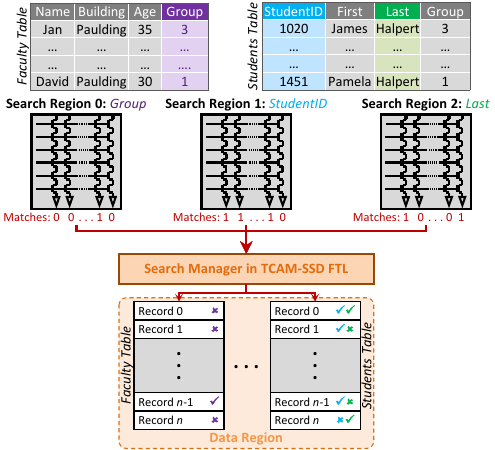}
	\caption{\sg{Search/data region mapping of database tables.}}
	\label{fig:db-search-regions}
\end{figure}

\sg{When the database is too large to fit in memory, TCAM-SSD can accelerate a transaction
by performing a bulk associative search across an entire column of the database.
Figure~\ref{fig:db-search-regions} shows an example of two tables (adapted from Aho et al.~\cite{Aho1992}): one containing records of faculty members, and another containing records of students.
Instead of using dedicated hash or tree data structures to index a column,
TCAM-SSD creates a search region for each indexable column,
where the region consists of the data stored for each record under that column.
For example, in the Faculty table, if the database management system (DBMS) wants to index the \textbf{Group} column,
the host reads and transposes the column's contents, and then calls \tSG{the TCAM-SSD \textit{Allocate/Append} command}
to write the transposed values into a newly-allocated search region.\footnote{\sg{Note that with computational SSD support, we could perform the transposition in the SSD instead of at the host, eliminating CPU--FE movement to generate the index.}}
Now, whenever a transaction wants to search for records using the \textbf{Group} column,
the DBMS can simply call the TCAM-SSD \textit{Search} command,
which performs a single-cycle parallel lookup across the entire index,
and copies any matching records from the data region (which in this case
is the already-stored database table) to the host buffer.}

\sg{With TCAM-SSD, the DBMS can easily create multiple indexes, with each getting its own search region.
As shown in Figure~\ref{fig:db-search-regions}, our example Student table
has indexes for both \textbf{StudentID} and \textbf{Last}.
Both of these search regions' link table entries point to the same data region,
avoiding the need to replicate the database tables.
TCAM-SSD can also support meta-index generation
(e.g., it can store an index with the length of each last name).}

\paragraph{Methodology}
We use \sg{the TPC-C database}~\cite{TPC-C} to evaluate how TCAM-SSD optimizes an OLTP system.
We \sg{generate a trace of \SI{1}{\mega\nothing} transactions 
for an OLTP workload using the DBx1000 DBMS}~\cite{Yu2014}, 
\sg{where the queries within the transactions use} a mixture of \revRW{indexes}.
\rw{We scale the number of \sg{records} in the table by 100, resulting in \SI{3}{\mega\nothing} entries.}
\sg{In our setup, the database is stored in the SSD, mimicking a setup where the
database is too large to fit entirely in memory.
For the baseline database running on a conventional SSD, all indexes are
stored in main memory.}

\sg{Our workload uses one secondary index (LastName), which the baseline system organizes as a hash
index. This leads to a number of collisions, where each collision needs to retrieve multiple
pages of data from the SSD to check for exact matches.
TPC-C records include a warehouse ID, but because the first query in each transaction limits
the search to only a single warehouse ID, we choose TCAM-SSD's search regions for the database
to correspond to a single warehouse.}

\begin{figure}[h]
    \vspace{-10pt}
    \subfloat[\revRW{\sg{CDF} of queries}]{\includegraphics[width=0.35\textwidth]{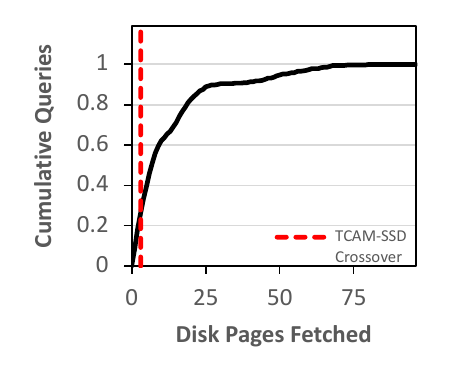}
    \label{fig:db-tpcc-cdf}
    }
    \subfloat[\revRW{\sg{CDF} of latency}]{\includegraphics[width=0.35\textwidth]{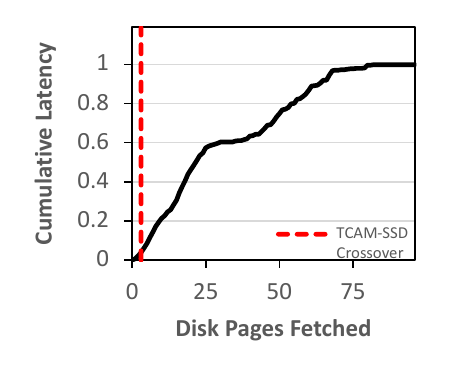}
    \label{fig:db-tpcc-lat}
    }
    \caption{\sg{Cumulative distribution function (CDF) showing which queries are accelerated by TCAM-SSD.]}}
    \label{fig:tpc-c}
\end{figure}

\paragraph{Results} %
We observe that over the course of the workload, TCAM-SSD achieves a speedup of \urevRW{60.9\%} over DBx1000.
\sg{To understand the source of this speedup, we examine which queries run faster on TCAM-SSD, and which queries run faster on DBx1000.}
We determine that TCAM-SSD is faster than DBx1000 whenever a query needs to retrieve more than 3~pages from the SSD.
Figure~\ref{fig:db-tpcc-cdf} shows the cumulative distribution function (CDF) of the queries in the workload,
as a function of the fetched page count for each query.
We observe from the figure that 73.5\% of queries exceed the 3-page threshold (i.e., run faster on TCAM-SSD than on DBx1000).
However, the greater the number of pages is per query, the greater the benefit that TCAM-SSD provides.
Therefore, we plot the CDF of latency in Figure~\ref{fig:db-tpcc-lat}, and observe that TCAM-SSD improves the latency for queries that take up \rw{95.8\%} of the total workload latency.%

\sg{TCAM-SSD reduces both CPU--FE and FE--BE data movement, by \rw{92.3\%} and \rw{77.0\%}, respectively, compared to DBx1000.
This reduction, due to TCAM-SSD's ability to execute the search directly inside the NAND flash array, 
contributes to its latency reduction while reducing the energy spent on disk I/O.
TCAM-SSD's overheads are small, requiring only 23 flash blocks ($<0.01\%$ of the SSD capacity) to store the search regions,
and only \tRW{\SI{2.5}{\kilo\byte}} of firmware DRAM to store the link tables.
We conclude that TCAM-SSD efficiently enables significant savings for our OLTP workload.}

\subsection{Database Analytics}
\label{sec:db:olap}
\label{ssec:db-olap}

\sg{An \emph{online analytics processing} (OLAP) workload consists of complex queries that scan entire columns of data from a table and aggregate the scanned information.
Unlike an OLTP workload, an OLAP workload does not make \tRW{efficient use of an index},
as the index is designed to avoid a column-wide scan to look up a specific column attribute,
while the OLAP workload wants to scan all attributes in the column.
As a result, OLAP workloads often incur significant SSD I/O whenever a database table does not fit in memory.}

\sg{TCAM-SSD can substantially mitigate the overheads of SSD I/O incurred during OLAP scan operations.
Recall from Section~\ref{sec:db:oltp} that TCAM-SSD implements an indexed column as a search region,
and performs highly-parallelized search across the entire column to locate matching records.
This is effectively a fast parallel scan operation, and we can reuse this same format and operation for OLAP scans.
Using our example from Figure~\ref{fig:db-search-regions}, 
a simple OLAP query may want to aggregate a list of all last names beginning with the letter H.
TCAM-SSD can quickly scan the column and return a list of all records matching the query,
generating CPU--FE data movement only for the relevant records
(as opposed to having to send \emph{all} records to the host for a conventional SSD).}

\sg{Note that in a conventional system, we often use different data organizations for
OLTP (row-major table storage) and OLAP (column-major table storage) to reduce I/O traffic.
However, with TCAM-SSD, we can use a single data organization for both while reducing I/O traffic even further.}
\tRW{Additionally, TCAM-SSD can generate a single search region for fused keys (i.e., a concatenation of two or more columns),
further reducing scan latency.}

\paragraph{Methodology}
\sg{We use TPC-H~\cite{TPC-H}, a business analytics workload, \rw{and populate the database using dbgen~\cite{dbgen}}.
With a scale factor of 100, the resulting database has a size of \SI{115}{\giga\byte}.}
\sg{We evaluate} two analytic queries examined in prior work~\cite{Gu2016, Woods2014}, which are modified versions of TPC-H queries \sg{that scan one \SI{74}{\giga\byte} table from the database}.
\sg{Query~2 performs additional filter operations compared to Query~1, and we use the fused key optimization in TCAM-SSD to efficiently support these filters.}

\paragraph{Results}
For the database that we generate, TCAM-SSD \tRW{on average} speeds up the scan operation for Query~1 by \urevRW{18.3}$\times$, and Query~2 by \urevRW{17.1}$\times$,
compared to a baseline database scan operation.
TCAM-SSD's improvements are a result of reducing both CPU--FE and FE--BE data movement (see Section~\ref{sec:framework:datamovement}).
For Query~1, the baseline must send the entire table's data to the host CPU,
requiring \SI{4.9}{\mega\nothing} read operations that generate \SI{74}{\giga\byte} of \sg{CPU--FE} \sg{\emph{and} \SI{74}{\giga\byte} of FE--BE} movement.
In comparison, TCAM-SSD requires only \SI{4.6}{\kilo\nothing} \emph{SRCH} chip commands (which generate \rw{no CPU--FE \sg{movement} and \SI{71.5}{\mega\byte} of FE--BE movement),
and only \SI{240.0}{\kilo\nothing} read operations for the matching data (\SI{3.7}{\giga\byte} of CPU--FE and \SI{3.7}{\giga\byte} of FE--BE movement).\footnote{\sg{While computational SSDs can achieve similar CPU--FE savings for analytics~\cite{lagrange.icpe2020, samynathan.adms2019}, they would still generate \SI{74}{\giga\byte} of FE--BE movement, saturating internal bandwidth and increasing energy.}}
Query~2's additional filter operations increase \emph{SRCH} command count to \SI{18.3}{\kilo\nothing}.
This increases FE--BE data movement to \rw{\SI{286.1}{\mega\byte}} (as more match vectors need to be transmitted to the firmware), 
but keeps CPU--FE data movement at only \SI{3.7}{\giga\byte}.
To enable these benefits, TCAM-SSD requires only 4578 NAND flash blocks (1.7\% of the SSD capacity) and \SI{0.2}{\mega\byte} of DRAM storage for the link table.}

\sg{Because an OLAP query typically returns many records, the location of each matching record has a significant impact on performance.
For example, if a query matches 10 records, these records could all be in the same NAND flash page, or they could each reside in different pages.
To explore the impacts of different database layouts, we analytically sweep two parameters for both of our queries:
\emph{selectivity} is the fraction of records in the database that match a query, and
\emph{locality} captures how likely the records are to share a page.
For example, a locality of 0\% means that we need an SSD read operation \emph{for every matching record},
while a locality of 100\% means that we assume all matching records are stored back-to-back and require the minimal number of SSD reads.
\tRW{By evaluating combinations of locality and selectivity, we aim to remain DBMS-agnostic, and demonstrate how TCAM-SSD can affect the performance of different data layouts}
}

\begin{figure}[h]
    \centering
    \vspace{-15pt}
    \subfloat[\revRW{Query 1 \sg{(y-axis is log scale)}}]{
        \includegraphics[width=0.4\textwidth, trim = 0 10 0 0, clip]{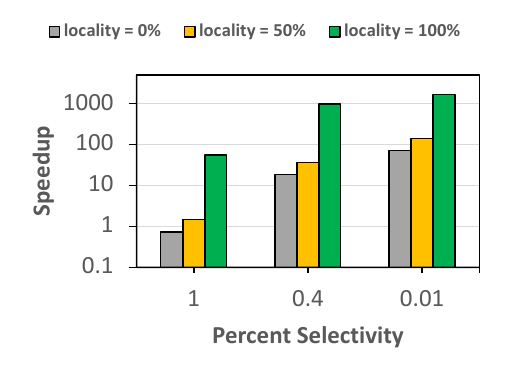}
        \label{fig:db-Query1}
    }%
    \subfloat[\revRW{Query 2 \sg{(y-axis is log scale)}}]{
        \includegraphics[width=0.4\textwidth, trim = 0 10 0 0, clip]{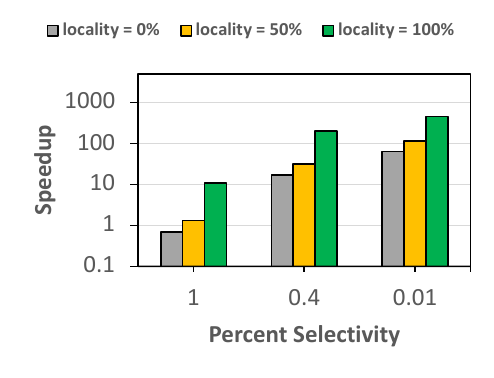}
        \label{fig:db-Query2}
    }%
    \caption{\revRW{Speedup for analytical queries with TCAM-SSD, normalized to scan \sg{using a conventional SSD}.}}
    \label{fig:db-olap}
\end{figure}

\sg{Figure~\ref{fig:db-olap} shows the impact of selectivity and locality on performance.
We make four observations from the figure.
First, TCAM-SSD's speedups over the baseline range from \urevRW{0.74}$\times$ (1\% selectivity, 0\% locality) to \urevRW{1637.0}$\times$ (0.01\% selectivity, 100\% locality), with an average speedup of \urevRW{113.5}$\times$ across the sweep.
Second, TCAM-SSD's benefits increase as the selectivity decreases.
Third, TCAM-SSD's benefits are smaller for Query~2 than for Query~1.
This is because Query~2 contains additional conditions for the scan.
However, TCAM-SSD still preserves much of the speedup by taking advantage of
its fused key optimization to reduce the impact of these additional conditions.
Note that for our synthesized database, the selectivity and locality for both queries is 0.04\% and 0.0\%, respectively.
We conclude that TCAM-SSD can substantially improve the performance of analytics,
and that its benefits increase with match sparsity and with worsening locality
(two common effects as the dataset size increases).}

%% file: sections/graph.tex
\section{Graph Processing Using TCAM-SSD}
\label{sec:graph}

\rw{\sg{Graph} processing is employed today across a wide variety of domains, ranging from social media networks~\cite{Yang2011}, to roads and geographical data~\cite{Leskovec2009}, to the connectivity of the Internet~\cite{Web}.
\sg{A graph processing framework~\cite{Pearce2010, Zheng2015} typically initiates analytics by preprocessing the graph from the dataset's generic input format (e.g., an edge array~\cite{Roy2013, Gonzalez2012}) into a framework-specific format (along with applying other optimizations, e.g., \cite{Maass2017, Zhu2015, Vora2016, Zhang2015}).}}
\rw{\sg{For} large networks, graph frameworks use \rw{meta}data structures such as \sg{in-memory indexes} that allow the system to quickly locate the edges of interest \sg{(which are stored in the SSD).}}

\begin{figure}[!h]
    \vspace{-10pt}
    \subfloat[Conventional adjacency list]{
        \includegraphics[width=0.3\columnwidth]{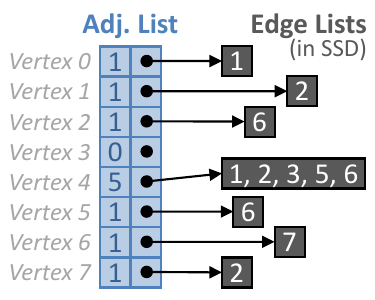}
        \label{fig:graph-adjlist}
    }
    \qquad%
    \subfloat[TCAM-SSD index]{
        \includegraphics[width=0.3\columnwidth]{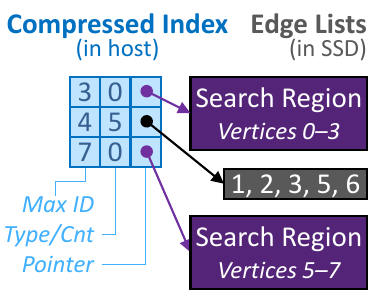}
        \label{fig:graph-spindex}
    }
    \caption{Comparison of graph index structures.}
    \label{fig:graph-overview}
\end{figure}

\sg{While many different metadata structures exist for the index~\cite{Zhu2015}, we focus on \emph{adjacency lists}, which form the basis of many graph analytics data structures.}
\sg{For each vertex, the adjacency list stores a count of the number of outgoing edges,
and a pointer to the first edge belonging to the vertex in the edge list.}
For the sake of simplicity, we describe how the index can be created for out edges; however, \sg{the} process can be repeated for in edges.
Figure~\ref{fig:graph-adjlist} shows an example 
adjacency list \sg{for eight vertices}. 

\rw{Notably, as \sg{graph datasets grow} in size, 
the size of the index grows, requiring additional memory to store both the index and edges.
Once the index exceeds the capacity of memory, systems that process large graphs experience severe performance degradation~\cite{Jun2018}, requiring the system to migrate both index and graph data between memory and SSD.
}
\rw{TCAM-SSD \sg{can optimize SSD I/O for large graphs by eliminating the conventional adjacency list and reducing the multi-step edge retrieval process into a single Search command}.}

\rw{%
\sg{A naive TCAM-SSD graph format could simply}
forgo an in-memory index, and instead \sg{perform bulk parallel search directly on the edge array} to find corresponding edges based on either source or destination \sg{vertices.}
\sg{This, however, is highly inefficient, because while TCAM-SSD's Search operation can search millions of edge entries simultaneously (accounting for maximum parallelism), large graphs can contain \emph{billions or even trillions} of edges.
This would require TCAM-SSD to perform thousands of back-to-back Search operations, and could stall unrelated I/O operations by tying up all of the SSD's channels.}
Instead, TCAM-SSD uses a compressed \sg{host-side} in-memory structure to reduce 
metadata overheads.}

\paragraph{Compressing the Index}
\rw{Graphs often follow a power law distribution~\cite{Page1999, Brin1998, Zheng2015, Chen2010}, where few vertices have \sg{a large degree (i.e., count of connected edges)}, while many vertices have \sg{a single-digit degree}.
\sg{If we were to allocate a dedicated search region to each vertex, most of the regions} 
would be \sg{highly} underutilized.
\sg{For the common case where} the out-degree of the vertex is less then the number of bitlines in the block, a Search operation will still \sg{check all of the empty bitlines in the block, wasting both energy} and area.
\sg{We propose two optimizations to reduce the underutilization,
resulting in the compressed index shown in Figure~\ref{fig:graph-spindex}.}}

First, \sg{multiple vertices with a small out-degree, and with consecutive IDs,} can be compressed into a single search region.
\sg{Our index stores a single entry for the search region, and stores the highest vertex ID in the Max ID column, %
along with a search region pointer.
To access the correct search region for a vertex, the graph framework performs a binary search over the sorted Max ID field.
For example, in Figure~\ref{fig:graph-spindex}, vertices 0--3 are compressed into a single search region, and vertices 5--7 are compressed into a second search region.
If the framework wants to locate the region for vertex~6, it uses the entry with a Max ID of 7, since 6 falls between 4 and 7.}

Second, certain vertices may contain substantially more edges than the number of \sg{bitlines} per block.
\sg{A TCAM-based search for such a vertex could tie up multiple levels of internal SSD parallelism,} and may return matches to most of its edges, \sg{resulting in little to no reduction in read count}. 
\sg{For such vertices, we do not store their edges in a search region.
Instead, we store an edge list in the data region, and keep a direct pointer to the edge list in the index (along with an edge count).
We set the edge count to 0 for search region entries, to distinguish it from an edge list entry.\footnote{\rw{This structure is inspired by prior work on sparse data structures~\cite{Vuduc03}, including those of graph analytics, and may be applicable \sg{for other} applications \sg{(e.g., sparse linear algebra)}.}}
Figure~\ref{fig:graph-spindex} shows how vertex 4 has a direct pointer to its edge list in our index.}

\begin{figure}[!h]
	\centering
	\includegraphics[width=.9\columnwidth]{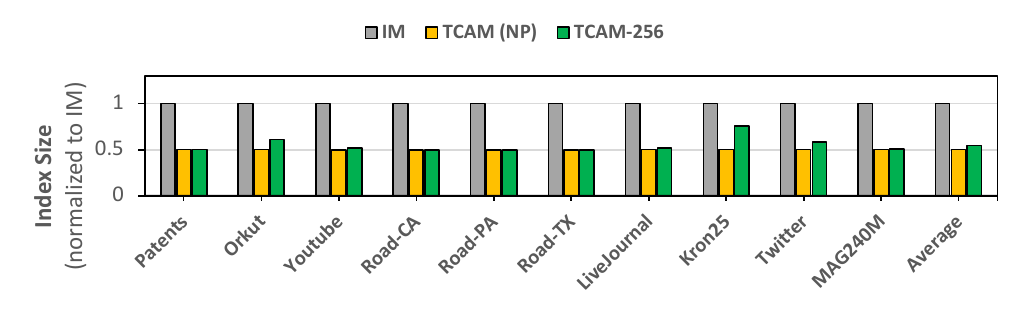}
	\caption{\revRW{Normalized graph index overhead.}}
	\label{fig:graph-index}
\end{figure}

\sg{TCAM-SSD constructs the compressed index during preprocessing, by performing a count sort (similar to the initial step of many graph preprocessing algorithms~\cite{Malicevic2017, Roy2015}).
Figure~\ref{fig:graph-index} shows the reduction in memory footprint across a variety of synthetic and real-world graphs (Table~\ref{tab:graphs}), for two TCAM-SSD configurations:
(1)~TCAM (NP), a basic version of our index without the large vertex optimization; and
(2)~TCAM-256, which uses a data region pointer for vertices with more than 256 edges.}
On average, TCAM-SSD reduces in-memory overheads by \tRW{47.5}\%, compared to a baseline index with a \SI{4}{\byte} pointer and \SI{4}{\byte} of metadata (e.g., vertex weight) \sg{per entry}.

\begin{table}[!h]
\vspace{-2pt}
\setlength\tabcolsep{2.5pt}
\caption{Evaluated graphs.}
\label{tab:graphs}
\centering
    \small
    \begin{tabular}{|c|c|c||c|c|c|}
        \hline
        \textbf{Graph} & \textbf{Nodes} & \textbf{Edges} & \textbf{Graph} & \textbf{Nodes} & \textbf{Edges} \\
        \hline
        Patents~\cite{Leskovec2005} & 3.7M & 16.5M & Orkut~\cite{Yang2012} & 3M & 117M \\
        \hline
        Road-CA~\cite{Leskovec2009} & 1.9M & 2.7M & Youtube~\cite{Yang2012} & 1.1M & 3M \\
        \hline
        Road-PA~\cite{Leskovec2009} & 1.1M & 1.5M & LiveJournal~\cite{Backstrom2006} & 4.8M & 69M \\
        \hline
        Road-TX~\cite{Leskovec2009} & 1.3M & 1.9M & Kron25 & 33.5M & 1B \\
        \hline
        Twitter~\cite{Yang2011} & 17M & 1.5B & \revRW{Mag240}~\cite{Wang2020} & \revRW{121.7M} & \revRW{1.3B}\\
        \hline
    \end{tabular}
\end{table}

\paragraph{Formulating Graph Analytics as a Search Problem}
\rw{To retrieve edge data \sg{for a vertex} from TCAM-SSD, the application \sg{starts by searching for the corresponding in-memory index entry.}}
\sg{For example, to retrieve data for vertex~2 in Figure~\ref{fig:graph-spindex}}, TCAM-SSD performs a binary search of the index structure \sg{in the following order: Max ID = 4, Max ID = 3 (which is the correct entry)}.
Once the correct entry is found, the system then issues either a search or read \sg{based} on the \sg{entry's edge list count.}
If the \sg{count is non-zero,} TCAM-SSD 
\sg{reads the edge list.}
If the \sg{count is 0, TCAM-SSD searches a search region}.
\rw{Within a search region, \sg{the bitline holds a (\emph{src}, \emph{dst}) tuple.}
When vertex~2 is used as \emph{src}, \sg{TCAM-SSD finds all matches in the search region, and accesses the corresponding data region to read out and return the out-edges.}}

\paragraph{Methodology}
To evaluate TCAM-SSD, we use a vertex access traversal trace for the SSSP algorithm.
We examine four different cases:
(1)~in-memory index (IM), where the LBA for each edge list is stored in memory;
(2)~out of memory (OOM), where both the edge list and index are on disk;
(3)~TCAM (NP); and
(4)~TCAM-256.
\rw{For TCAM (NP) and TCAM-256, we \sg{assume that every index access is a DRAM row miss.}}
We conservatively assume 0\% locality, and explore the interplay between search and data pointers in the index.

\begin{figure}[!ht]
	\centering
	\includegraphics[width=.9\textwidth]{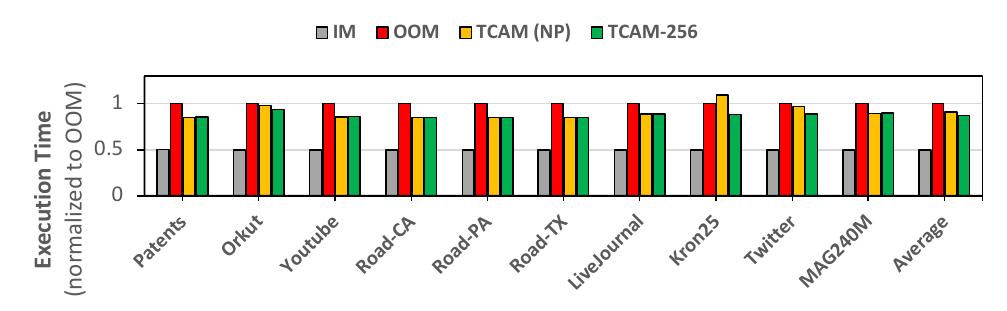}
	\caption{\revRW{Execution time for SSSP.}}
	\label{fig:graph-time}
\end{figure}

\paragraph{Results}
Figure~\ref{fig:graph-time} shows the execution time for SSSP vertex traversals on our four configurations, normalized to IM.
We make \tSG{four} observations from the figure.
First, OOM incurs a 99\% overhead over IM, averaged across all networks, indicating the storage access cost of large graphs in conventional systems.
Second, TCAM (NP) performs \tRW{10.2}\% better than OOM on average, as it avoids data movement costs between the CPU and the SSD.
\tRW{
Third, \tSG{while TCAM (NP) improves performance for most datasets, we do see performance degradation for Kron25}.
We determine that this is due to the overheads incurred from moving and decoding the match vector for vertices with \tSG{a} high degree. 
\tSG{Fourth, because TCAM-256 uses our optimized data structure, it} can further improve the performance for large graphs with vertices with high degrees with an average improvement of {14.5}\% and {4.3}\% over OOM and TCAM (NP), respectively.
For Kron25, TCAM-256 results in a 24.2\% speedup compared to TCAM (NP), \tSG{showing that our optimization overcomes the performance overheads of TCAM-SSD for high-degree vertices}.}

For the synthetic Kron25 graph, the search region uses ~\tRW{8200 blocks (3.1\% of SSD capacity)}
and the link table overhead is \SI{66}{\mega\byte}.
For the real-world Twitter graph, the search region overhead is \tRW{3.8\%} with a link table overhead of \SI{50.9}{\mega\byte}.
Although Twitter contains more edges than kron25 (Table~\ref{tab:graphs}), kron25 has more high-degree vertices, necessitating more link table entries.

%% file: sections/related.tex
\section{Related Work}
\label{sec:related}

\sg{To our knowledge, TCAM-SSD is the first end-to-end framework for in-NAND-flash associative \tRW{searching}.
We discuss closely-related works below.}

\paragraph{In-Storage Computation.}
Parabit~\cite{Gao2021}, Flash-Cosmos~\cite{Park2022}, and GP3D~\cite{Shim2022} perform processing using NAND flash memory. TCAM-SSD avoids shortcomings of these works by enabling general-purpose associative computing instead of bulk bitwise or domain-specific solutions.
Computational SSDs (i.e., smart SSDs) exploit the internal microcontroller~\cite{SmartSSDsas, Kwak2020, Do2013, Kang2013, Koo2017, samsung.smartssd.gen2.pressrelease} or nearby FPGAs~\cite{Daisy2020, Salamat2021, samsung.smartssd.website} to perform processing-near-memory.
Other approaches~\cite{Mailthody2019, Bae2013, Cho2013, Jun2015, Kim2011, samsung.smartssd.gen2.pressrelease} place additional hardware throughout the SSD hierarchy to enable in-storage computation. %
Some in-storage computing platforms target query processing~\cite{Park2021sql, Seshadri2014, Xu2020, Hu2022} or key--value interfaces~\cite{Jin2017, Xu2016, Im2020}.
TCAM-SSD is orthogonal to these computational SSD solutions, and can benefit from offloading certain operations to embedded compute elements.

SmartSSDs are an emerging technology that attempt to exploit the increasingly powerful microcontroller on SSDs~\cite{SmartSSDsas, Kwak2020, Do2013, Kang2013, Koo2017}.
These systems are sometimes augmented with field-programmable gate arrays (FPGAs) to further enhance their compute capabilities~\cite{Daisy2020, Salamat2021}.
Similarly, other techniques have been proposed to execute query processing on the SSD~\cite{Park2021sql, Seshadri2014, Xu2020}.
Other approaches~\cite{Mailthody2019, Bae2013, Cho2013, Jun2015, Kim2011} have explored placing additional hardware throughout the SSD hierarchy to enable in-storage computation. %
Another class of systems~\cite{Jin2017, Xu2016, Im2020} integrate key-value interfaces onto the SSD to further accelerate data retrieval.
TCAM-SSD is compatible with many of the above systems; notably, TCAM-SSD can benefit from offloading portions of the computation onto the SSD microcontroller.

\paragraph{\tRW{Key--Value Stores.}}
\tRW{Key--value stores (KVS)~\cite{DeCandia2007, Debnath2010, Atikoglu2012} are a software-defined construct that map a predefined key to a cluster of data.
Using an input key, \tSG{a KVS} can leverage the associative search operations of TCAMs to retrieve corresponding clusters of data in O(1) time.
Although associative memories can be used to implement \tSG{a key--value store, a KVS does not} require explicit use of TCAMs~\cite{Jin2017, Kang2019, Kaiyrakhmet2019}.}

\paragraph{Associative Computing and Content-Addressable \tSG{Parallel} Processors.}
\tRW{Associative computing (\tSG{i.e., associative} processing)~\cite{Potter2012} broadly refers to \tSG{a} computing paradigm in which associative operations are used to locate and compute over data in parallel.}
Associative computing has been explored in a variety of contexts~\cite{Foster1976, Potter1994, Sayre1976, Pagiamtzis2006, Yavits2021},
\tSG{and often (but not always) relies} on associative \tSG{memory structures}~\cite{Foster1976, Guo2013, Caminal2021, Caminal2022}.
\tSG{Two such examples are CAPE~\cite{Caminal2021}, which accelerates associative computing in SRAM, and Castle~\cite{Caminal2022}, which} extends upon CAPE to accelerate databases.
Other associative computing frameworks~\cite{Zha2020, Guo2011, Guo2013} employ emerging NVM technologies, limiting their near-term applicability.
In contrast, TCAM-SSD implements associative computing \sg{using conventional} NAND flash memory, which provides high storage density in a mature technology.

\tSG{There are multiple approaches and definitions of associative computing.}
One example framework, ASC~\cite{Potter1994}, provides \tSG{a mechanism to convert} conventional RAM-based algorithms into associative computing algorithms by \tSG{organizing} data into two-dimensional tables.
To \tSG{perform computation}, bulk associative search operations are then used to select the corresponding entries and retrieve the matching data element for addition computation.
\tRW{\tSG{A second example is the} content-addressable parallel processor (CAPP)~\cite{Foster1976}, which \tSG{consists of three formal} requirements: 
(1)~\tSG{it stores} data in \tSG{a vector format},
(2)~\tSG{it can} compare key against all vector elements in parallel,
(3)~\tSG{it can} update all matching elements in bulk with a new value.
Although the TCAM-SSD framework broadly fits into the associative computing paradigm, as it relies on associative search operations to find and retrieve data, it is not a CAPP, as it does not natively support updating all matching elements in parallel. \tSG{However, we describe how TCAM-SSD can be extended to support this using its associative update mode}
(Section~\ref{sec:framework:software}).}

\paragraph{Database Acceleration.}
Recent works propose to use emerging NVM technologies for analog \textit{in situ} SQL-style operations.
Sun et al.\ propose a storage scheme for in-memory databases to map tuples onto crossbars~\cite{Sun2017, Wang2018}, while others~\cite{Li2020} utilize ReRAM-based content addressable memories (ReCAMs) to provide \textit{in situ} support for fundamental SQL database operations, including sort, join, and selection.
\sg{Other recent} work has explored the use of Optane~\cite{Hady2017, Wu2019b, Shanbhag2020} and CXL~\cite{Ahn2022} to extend the memory capacity available to the DBMS,
while others seek to mitigate the overheads of pointer chasing during data retrieval~\cite{Kocherber2013}.
FPGAs have also shown promise in accelerating database workloads~\cite{Kim2019}.

\paragraph{Graph Workload Acceleration.}
Graph analytics has been widely explored through a variety of hardware and software frameworks~\cite{Maass2017, Liu2017}. %
PREGEL~\cite{Malewicz2010} introduces the ``think like a vertex'' framework, enabling a simple programming model for distributed graph processing.
GraphChi~\cite{Kyrola2012} and X-Stream~\cite{Roy2013} demonstrate the potential for graph processing without assumptions of memory size, by effectively utilizing secondary storage.
Ligra~\cite{Shun2013} proposes a high-performance shared memory machine framework that assumes the graph fits in a the memory system.
Recent works propose SSD optimizations for graph analytics~\cite{Zheng2015, Jun2018, Suzuki2021, Matam2019}
or dedicated accelerators for graph processing~\cite{Ham2016, Dadu2021, Ahn2015}. %

%% file: sections/conclusion.tex
\section{Conclusion}
\label{sec:conc}

We present TCAM-SSD, a framework for \tRW{in-SSD associative search using NAND flash memory}. 
With modest modifications to the NAND flash chips inside commodity solid-state drives (and with no modifications to the NAND flash array), we can enable highly-parallel ternary search operations.
Our framework includes a hardware primitive, firmware modifications, and a user interface, along with hardware optimizations and application-specific optimizations.
\tRW{We show that for three use cases, TCAM-SSD can provide notable performance and data movement improvements for large dataset processing.}

%% file: sections/sandiastatement.tex
This article has been authored by an employee of National Technology \& Engineering Solutions of Sandia, LLC under Contract No. DE-NA0003525 with the U.S. Department of Energy (DOE). 
The employee owns all right, title and interest in and to the article and is solely responsible for its contents. 
The United States Government retains and the publisher, by accepting the article for publication, acknowledges that the United States Government retains a non-exclusive, paid-up, irrevocable, world-wide license to publish or reproduce the published form of this article or allow others to do so, for United States Government purposes. 
The DOE will provide public access to these results of federally sponsored research in accordance with the DOE Public Access Plan \url{https://www.energy.gov/downloads/doe-public-access-plan}. 
This paper describes objective technical results and analysis. 
Any subjective views or opinions that might be expressed in the paper do not necessarily represent the views of the U.S. Department of Energy or the United States Government.